\documentclass[twocolumn]{aastex63}

\def\a{\alpha}
\def\b{\beta}
\def\g{\gamma}
\def\y{\gamma}
\def\d{\delta}
\def\D{\Delta}
\def\e{\epsilon}
\def\i{\iota}
\def\k{\kappa}
\def\l{\lambda}
\def\u{\mu}
\def\v{\nu}
\def\s{\sigma}
\def\t{\tau}

\def\x{\xi}

\def\n{\nabla}
\def\w{\omega}
\def\z{\zeta}

\def\p{\phi}

\def\L{\Lambda}

\def\O{\Omega}
\usepackage{lipsum}

\shorttitle{\textsc{X-ray and Radio Upper Limits to Gaia BHs}}
\shortauthors{Rodriguez et al.}
\graphicspath{{./}{figures/}}

\usepackage{float}
\usepackage{appendix}
\usepackage{xcolor}
\usepackage{amsmath}

\begin{document}

\title{No X-Rays or Radio from the Nearest Black Holes and Implications for Future Searches}

\correspondingauthor{Antonio C. Rodriguez}
\author[0000-0003-4189-9668]{Antonio C. Rodriguez}\thanks{E-mail: acrodrig@caltech.edu}
\affiliation{Department of Astronomy, California Institute of Technology, 
1200 E. California Blvd, Pasadena, CA, 91125, USA}

\author{Yvette Cendes}
\affiliation{Center for Astrophysics \textbar  Harvard \& Smithsonian, 60 Garden Street, Cambridge, MA, 02138, USA}
\affiliation{Department of Physics, University of Oregon, Eugene, OR 97403, USA}

\author{Kareem El-Badry}
\affiliation{Department of Astronomy, California Institute of Technology, 
1200 E. California Blvd, Pasadena, CA, 91125, USA}

\author{Edo Berger}
\affiliation{Center for Astrophysics \textbar  Harvard \& Smithsonian, 60 Garden Street, Cambridge, MA, 02138, USA}

\begin{abstract}
Astrometry from the {\it Gaia} mission was recently used to discover the two nearest known stellar-mass black holes (BHs), Gaia BH1 and Gaia BH2. Both systems contain $\sim 1\,M_{\odot}$ stars in wide orbits ($a\approx$1.4 AU, 4.96 AU) around $\sim9\,M_{\odot}$ BHs. These objects are among the first stellar-mass BHs not discovered via X-rays or gravitational waves. The companion stars --- a solar-type main sequence star in Gaia BH1 and a low-luminosity red giant in Gaia BH2 --- are well within their Roche lobes.  However, the BHs are still expected to accrete stellar winds, leading to potentially detectable X-ray or radio emission. Here, we report observations of both systems with the {\it Chandra} X-ray Observatory, the VLA (for Gaia BH1) and MeerKAT (for Gaia BH2). We did not detect either system, leading to X-ray upper limits of $L_X < 10^{29.4}$ and $L_X < 10^{30.1}\,\rm erg\,s^{-1}$ and radio upper limits of $L_r < 10^{25.2}$ and $L_r < 10^{25.9}\,\rm erg\,s^{-1}$. For Gaia BH2, the non-detection implies that the the accretion rate near the horizon is much lower than the Bondi rate, consistent with  recent models for hot accretion flows. We discuss implications of these non-detections for broader BH searches, concluding that it is unlikely that isolated BHs will be detected via ISM accretion in the near future. We also calculate evolutionary models for the binaries' future evolution using Modules for Experiments in Stellar Astrophysics (MESA). We find that Gaia BH1 will be X-ray bright for 5--50 Myr when the star is a red giant, including 5 Myr of stable Roche lobe overflow. Since no symbiotic BH X-ray binaries are known, this implies either that fewer than $\sim 10^4$ Gaia BH1-like binaries exist in the Milky Way, or that they are common but have evaded detection, perhaps due to very long outburst recurrence timescales. 
\vspace{1cm}
\end{abstract}

\section{Introduction}
Understanding the full demographics of the stellar-mass black hole (BH) population provides key insights into stellar and galactic evolution. BHs are created by the deaths of some stars with initial masses $M_* \gtrsim 20M_\odot$. Precisely which stars form BHs, and which leave behind neutron stars or no remnant at all, is uncertain \citep[e.g.][]{OConnor2011, Sukhbold2016, Laplace2021}. The Milky Way has formed $\sim10^{11}$ stars in its lifetime, and the stellar initial mass function \citep[IMF; e.g.][]{1955salpeter} dictates that the number of massive stars that have formed, died, and left behind a BH stands at  $\sim 10^7 - 10^8$ \citep[e.g.][]{Sweeney2022}.

Most ($\gtrsim 70\%$) of these massive stars exist with a binary companion, with triple and higher order systems also being common \citep{2007kobulnicky,2012sana, 2017moe}. However, the binary fraction of BHs is unknown. Virtually all known or suspected stellar-mass BHs today are in close binaries, in which a stellar companion to a BH is close enough that the BH is accreting significant quantities of gas from it, and the accretion flow produces observable emission across the electromagnetic spectrum. $\sim$20 dynamically confirmed BHs exist in X-ray binaries,  $\sim$50 X-ray sources are suspected to contain a BH based on their X-ray properties \citep[e.g.][]{2006mcclintock, 2016corral}, and a few X-ray quiet binaries have been reported in which a BH is suspected on dynamical grounds. Just one isolated BH candidate has been discovered via microlensing \citep{2022sahu, 2022lam, 2022mroz}.

In X-ray bright systems, a BH accretes material from a close companion through stable Roche lobe overflow or stellar winds \citep[e.g.][]{2006mcclintock}. These systems are called X-ray binaries (XRBs), and are often placed into three distinct spectral/temporal states: 1) the high/soft state, where the system is X-ray bright ($L_X \sim L_\textrm{Edd}$) and dominated by thermal emission from the accretion disk, 2)  the low/hard state, where the system is less X-ray luminous ($L_X \sim 10^{-2} L_\textrm{Edd}$) and dominated by power-law emission, and the 3) quiescent state, where the system is very faint ($L_X \lesssim 10^{-3} L_\textrm{Edd}$) and still dominated by power-law emission \citep[e.g.][]{2006mcclintock}. All BH XRBs have been discovered in either a persistent high/soft state or through transient X-ray nova (outburst) events \citep[e.g.][]{2016corral}. X-ray novae are caused by a sudden increase of mass transfer onto the BH, which leads to a dramatic increase in X-ray luminosity from quiescence  \citep[e.g.][]{2006mcclintock}. All-sky X-ray monitors \citep[e.g. MAXI, Swift/BAT;][]{2009maxi, 2005swift} have been effective at discovering BH candidates from their X-ray novae across the entire Milky Way for those whose luminosities approach $L_\textrm{Edd}$, and out to a few kpc for those that reach $\sim 10^{-2}L_\textrm{Edd}$ \citep[e.g.][]{2016corral}. However, the recurrence timescale of these outbursts is under strong debate, and so the total number of BH XRBs in the Galaxy is still quite uncertain \citep{1996xraynovae, 2022maccarone_outburst, 2022mori_reply}.

In the last few years, a handful of BHs orbited by luminous stars have been discovered in wider orbits  \citep{2018giesers, 2022shenar, 2023bh1, 2023bh2}. These systems are still outnumbered by XRBs, but this is likely a consequence of the very different selection functions of X-ray and optical searches. The few wide systems discovered so far likely represent the tip of a substantial iceberg. In this paper, we focus on Gaia BH1 and BH2, the newest and nearest of these systems. Precision astrometry from the third data release of the \textit{Gaia} mission (DR3) enabled their discovery, and optical high-resolution spectroscopy confirmed their nature. Gaia BH1 and BH2 are systems with a BH in an orbit with a sun-like main-sequence star, and a red giant likely in its first ascent of the giant branch, respectively. These systems are unique in currently being the BH binaries with the longest known orbital periods (186 days, 1277 days), largest binary separations ($a=$1.4 AU, 4.96 AU), and also the closest to Earth (480 pc, 1.16 kpc). 

Since both sun-like stars and red giants have stellar winds \citep[e.g.][]{1958parker, 1966redgiant_winds}, we asked: can we see evidence of wind accretion in Gaia BH1 and BH2? In Section \ref{sec:data}, we describe our observations and calculate upper limits for both Gaia BH1 and BH2 based on X-ray data from \textit{Chandra}/ACIS-S, and radio data from the Very Large Array (VLA) and MeerKAT. In Section \ref{sec:analysis}, we show that under the assumption of Bondi-Hoyle-Littleton (BHL) accretion, we should have seen X-rays and radio from Gaia BH2. We argue that a lack thereof signals that radiatively inefficient accretion is responsible for reduced accretion rates and the subsequent lack of multiwavelength emission. Finally, in Section \ref{sec:searches}, we explore the prospects of detecting wind accretion onto BHs using rates and efficiencies assuming inefficient accretion flow, either through a red giant companion or from the interstellar medium (ISM). We show that surveys such as SRG/eROSITA and pointed observations from \textit{Chandra} are at best sensitive to 1) BHs accreting from $\sim100 R_\odot$ red giants and 2) BHs accreting from high density ($n \gtrsim 10^3 \;\textrm{cm}^{-3}$) H$_2$ regions while traveling at very low ($\lesssim 5$ km/s) speeds. Finally, we present MESA models for the future evolution of both systems and their expected X-ray luminosities. Based on these models and the lack of detections of symbiotic BH XRBs from all-sky surveys, we conclude that at most $\sim10^4$ systems similar to either Gaia BH1 or BH2 exist in the Milky Way, unless a substantial population of symbiotic BH XRBs have evaded detection so far.

\section{Data}
\label{sec:data}

\subsection{X-Ray}
We observed Gaia BH1 with the \textit{Chandra} X-ray Observatory (ObsID: 27524; PI: Rodriguez) using the Advanced CCD Imaging Spectrometer (ACIS; Garmire et al. 2003) on 31 October 2022 (UT) for a cumulative time of 21.89 ks (sum of two observations: 12.13 ks and 9.76 ks). The ACIS-S instrument was used in pointing mode, chosen over ACIS-I for its slight sensitivity advantage. The observations were taken about 9 days before apastron, when the separation between the BH and star was $\approx$2.01 AU.

We observed Gaia BH2 for 20ks  with {\it Chandra}  on 2023 January 25 (proposal ID 23208881; PI: El-Badry). We also used the ACIS-S configuration, with a spatial resolution of about 1 arcsec. The observations were timed to occur near the periastron passage, when the separation between the BH and the star was $\approx$ 2.47 AU. 

The X-ray images of both sources are shown in Figure \ref{fig:imgs}. We first ran the \texttt{chandra\_repro} tool to reprocess the observations; this creates a new bad pixel file and de-streaks the event file. Since the observation of Gaia BH1 was split into two, we then ran the \texttt{reproject\_obs} tool to merge both observations with respect to a common World Coordinate System (WCS). We then performed aperture photometry at the optical positions of Gaia BH1 and BH2 in the reprocessed event files using the \texttt{srcflux} tool. We detect no significant flux at the location of either system and obtain a background count rate of ($2.0 \times 10^{-4}, \;1.5 \times 10^{-4} \;\textrm{cts s}^{-1}$), for Gaia BH1 and BH2 respectively.

In order to convert to unabsorbed flux, we assume a power-law spectrum with index of 2, and calculate the Galactic hydrogen column density using the relation from \cite{Guver2009} and the value of $A_V$ from \cite{Green2019, 2022lallement} --- $A_V=0.93\pm 0.1, 0.62\pm 0.1$ for BH1, BH2, respectively. Using the \texttt{PIMMS} tool, we calculate a background (1$\sigma$) unabsorbed X-ray flux of $F_X = 3.6\times10^{-15} \textrm{erg s}^{-1}\textrm{cm}^{-2}$ (BH1) and $F_X = 2.6\times10^{-15} \textrm{erg s}^{-1}\textrm{cm}^{-2}$ (BH2) in the 0.5--7 keV energy range. With the \textit{Gaia} distances, we can calculate 3$\sigma$ upper limits on flux and luminosity, which we present in Table \ref{tab:data}.

\subsection{Radio}
We observed Gaia BH1 for 4 hr with the Very Large Array (VLA) in C band (4--8 GHz) in the ``C" configuration on 2022 November 27--28 (DDT 22B-294; PI: Cendes). At this time, Gaia BH1 was 17 days past  apastron, and the separation between the BH and star was $\approx$1.98 AU. Data was calibrated using the Common Astronomy Software Applications (CASA) software using standard flux and gain calibrators. We
measured the flux density using the imtool package within pwkit
\citep{2017williams} at the location of Gaia BH1. The RMS at the
source’s position is 3.4 $\mu$Jy, and we present flux and luminosity 3$\sigma$ upper limits in Table \ref{tab:data}.

We observed Gaia BH2 for 4 hr with the MeerKAT radio telescope
in L band (0.86--1.71 GHz) on 2023 January 13 (DDT-20230103-YC01; PI: Cendes), when the separation between the BH and the star
was $\approx$ 2.54 AU. We used the flux calibrator J1939-6342 and the gain
calibrator J1424-4913, and used the calibrated images obtained via
the SARAO Science Data Processor (SDP)4 for our analysis.  We
measured the flux density using the imtool package within pwkit
\citep{2017williams} at the location of Gaia BH2. The RMS at the
source’s position is 17 $\mu$Jy, and we present flux and luminosity 3$\sigma$ upper limits in Table \ref{tab:data}.

\begin{figure*}
    \centering
    \includegraphics[width=\textwidth]{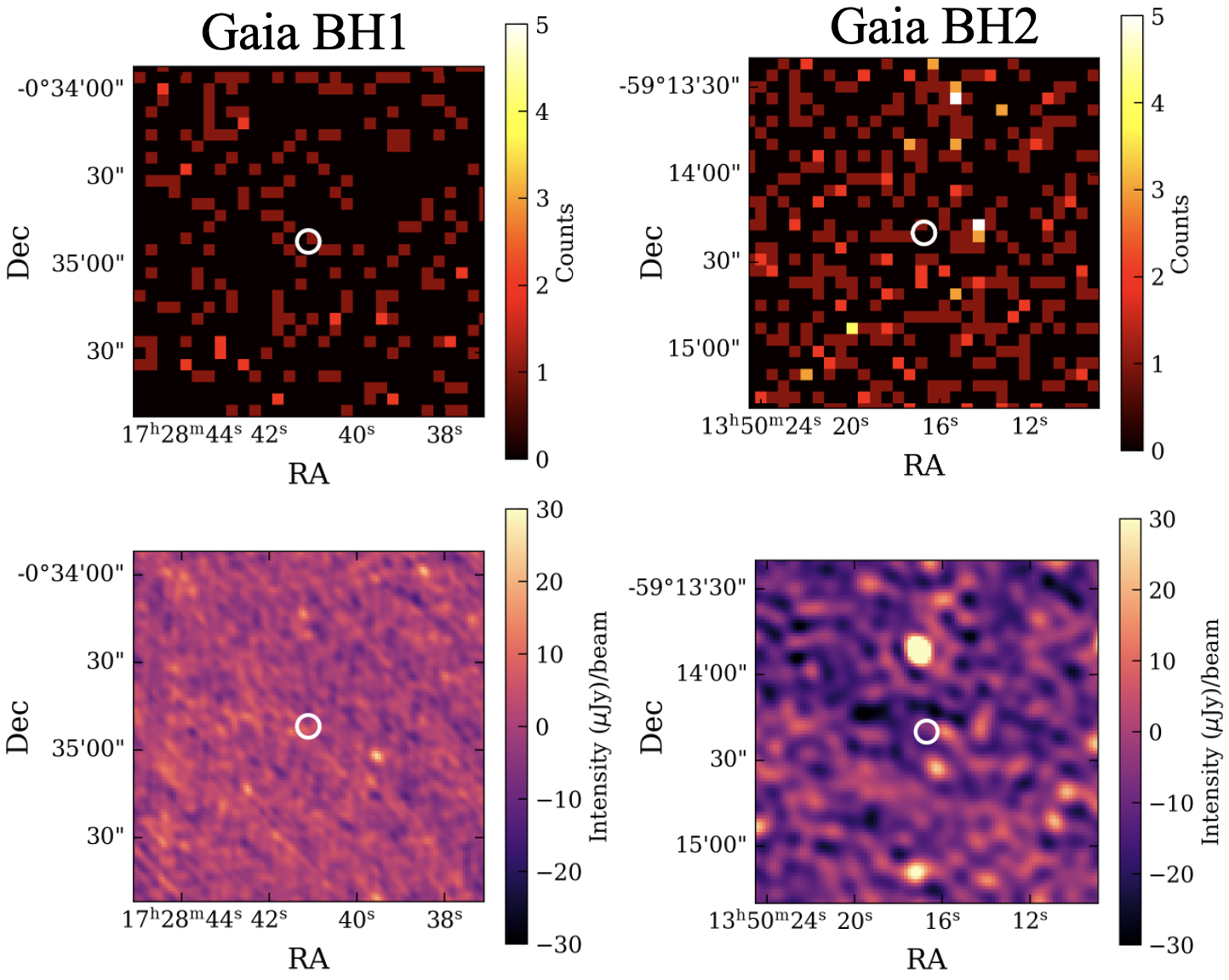}
    \caption{Images of Gaia BH1 (left panels) and Gaia BH2 (right panels) in the X-ray (upper panels) and the radio (lower panels). Both sources were observed for $\approx$ 20 ks with \textit{Chandra}/ACIS-S, corresponding to a flux limit of $\sim 5\times 10^{-15}$. Gaia BH1 was observed with the VLA for $\approx$4 hrs, and Gaia BH2 was observed with \textit{MeerKAT} for $\approx$4 hrs. No significant source of flux is detected at the position of Gaia BH1 or BH2 in either X-rays or radio.}
    \label{fig:imgs}
\end{figure*}

We show cutouts of all X-ray and radio images in Figure \ref{fig:imgs}. No significant source of flux is detected at the position of Gaia BH1 or BH2 in either X-rays or radio.

\begin{table*}[]
    \centering
    \caption{X-ray and Radio Upper Limits on Flux and Luminosity}
    \begin{tabular}{c|c|c|c|c|c}
         Object & Facility & Energy Range & BH--star separation & Flux Limit (3$\sigma$) & Luminosity Limit (3$\sigma$)\\
         \hline
         Gaia BH1 & Chandra/ACIS-S & 0.5--7 keV & 2.01 AU & $<1.1\times10^{-14} \textrm{erg s}^{-1}\textrm{cm}^{-2}$ & $< 2.8\times10^{29} \textrm{erg s}^{-1}$\\
         Gaia BH1 & VLA/C band & 4--8 GHz & 1.98 AU & $<$10.2 $\mu$Jy & $< 1.6\times10^{25} \textrm{erg s}^{-1}$\\
         \hline
         Gaia BH2 & Chandra/ACIS-S & 0.5--7 keV & 2.47 AU & $<7.8\times10^{-15} \textrm{erg s}^{-1}\textrm{cm}^{-2}$ & $< 1.2\times10^{30} \textrm{erg s}^{-1}$\\
         Gaia BH2 & MeerKAT/L band & 0.86--1.71 GHz & 2.54 AU & $<$51 $\mu$Jy & $< 7.8\times10^{25} \textrm{erg s}^{-1}$\\
    \end{tabular}
    \label{tab:data}
\end{table*}

\section{Theoretical predictions}
\label{sec:analysis}
\subsection{X-ray Estimates}
We first calculate the X-ray luminosity expected if the BHs accrete their companion stars' winds at the Bondi-Hoyle-Littleton (BHL) rate:

\begin{align}
    \dot{M}_\textrm{BHL} = \frac{4\pi G^2M_\textrm{BH}^2\rho}{(v^2 + c_s^2)^{3/2}}=\frac{G^2M_\textrm{BH}^2\dot{M}_\textrm{wind}}{v^4_\textrm{wind}d_\textrm{sep}^2}
    \label{eq:bhl}
\end{align}
where $d_\textrm{sep}$ is the separation between the star and BH (which varies as a function of position along the orbit in an elliptical orbit), $M_\textrm{BH}$ is the mass of the BH, $\rho$ is the density of accreted material, $\dot{M}_\textrm{wind}$ is the mass loss rate of the donor, $c_s$ is the sound speed, $v$ is the relative velocity between the BH and the accreted material, and $v_\textrm{wind}$ is the wind speed. We note that the second equality assumes the relative velocity between the BH and accreted material (i.e. the wind speed) greatly exceeds the sound speed. We assume that a fraction $\eta$ of the accreted rest mass is converted to X-rays, leading to an observable X-ray flux:
\begin{align}
    F_\textrm{X, BHL} = \frac{\eta \dot{M}_\textrm{BHL} c^2}{4\pi d^2} =  \frac{G^2c^2 M_\textrm{BH}^2}{4\pi d^2d_\textrm{sep}^2}\frac{\eta \dot{M}_\textrm{wind}}{v^4_\textrm{wind}}
    \label{eq:fx}
\end{align}
where $d$ is the distance to the system from Earth. We are left with an X-ray flux that depends on three unknown physical quantities: $\dot{M}_\textrm{wind}$, the mass loss rate of the donor star due to winds, $v_\textrm{wind}$, the wind speed, and $\eta$, the radiative efficiency of accretion. It is important to note that $\eta$ may vary with accretion rate, or with other properties of the accretion flow.

\begin{figure}
    \centering
    \includegraphics[width=0.48\textwidth]{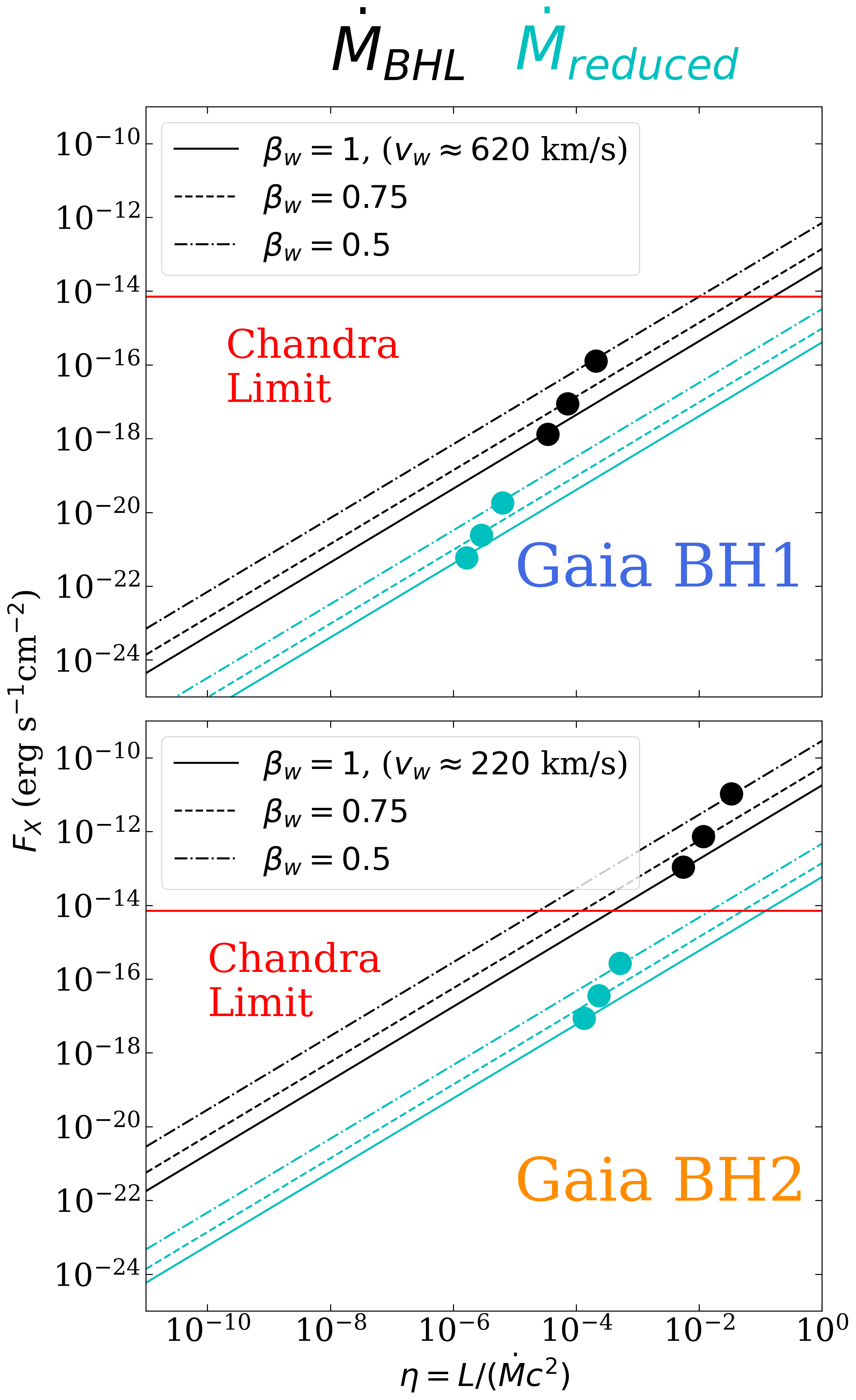}
    \caption{For all plausible wind speeds under the assumption of BHL accretion (black lines), \textit{Chandra} should have detected X-rays from Gaia BH1 if the accretion flow were radiatively efficient ($\eta \gtrsim 0.1$; top panel) and from Gaia BH2 if the accretion flow were radiatively \textit{in}efficient down to $\eta \gtrsim 10^{-4}$ (bottom panel). Black dots show expected efficiencies from models of hot accretion flows, but assuming the BHL accretion rate. Neither system is detected in X-rays, due to a combination of reduced accretion rates compared to the BHL assumption (cyan lines), and ensuing lower radiative efficiencies (cyan dots).}
    \label{fig:xray_est}
\end{figure}

The donor star in Gaia BH1 is a main-sequence Sun-like star (G dwarf), while the donor in Gaia  BH2 is a lower red giant ($R\sim 8\,R_{\odot}$). Since the donor in Gaia BH1 closely resembles the Sun, and abundance measurements point to it being $\gtrsim$ 4 Gyr old, we adopt a solar mass loss rate: $\dot{M}_\textrm{wind} \approx 2\times 10^{-14} M_\odot \textrm{yr}^{-1}$ \citep{1998wang_solarmassloss}. Gaia BH2, however, hosts a red giant donor star, which has a mass loss rate strongly dependent on stellar properties. We adopt a simple estimate for its mass loss rate from \cite{1975reimers}:

\begin{align}
    \dot{M}_\textrm{wind} = 4 \times 10^{-13}\beta_R \left(\frac{L_\star}{L_\odot} \right)\left(\frac{R_\star}{R_\odot} \right)\left(\frac{M_\star}{M_\odot} \right)^{-1} M_\odot \;\textrm{yr}^{-1}
    \label{eq:mdot_wind}
\end{align}
where $\beta_R$ is a scaling parameter for the mass loss rate. We set $\beta_R=0.1$, following empirical estimates for red giant branch stars \citep{1975reimers, 2016choi}. We approximate the wind speed as the escape velocity times a scaling parameter, $\beta_\textrm{wind}$:
\begin{align}
    v_\textrm{wind} = 600 \beta_\textrm{wind}\left(\frac{M_\star}{M_\odot} \right)^{1/2}\left(\frac{R_\star}{R_\odot} \right)^{-1/2} \textrm{km s}^{-1}
    \label{eq:v_wind}
\end{align}
We can then write a scaling relation for Equation \ref{eq:fx}, assuming fiducial values for Gaia BH1:
\begin{equation}
\label{eq:scale_bh1}
\begin{split}
F_{\rm X, BHL}	&=10^{-18}\,{\rm erg\,s^{-1}\,cm^{-2}} \left(\frac{\eta}{10^{-4}}\right)\left(\frac{M_{{\rm BH}}}{10\,M_{\odot}}\right)^{2}\times\\
&\left(\frac{d_{{\rm sep}}}{2.03\,{\rm AU}}\right)^{-2}\left(\frac{d}{480\,{\rm pc}}\right)^{-2}\times\\&\left(\frac{\dot{M}_\textrm{wind}}{2\times10^{-14} M_\odot \textrm{ yr}^{-1}}\right)\times \left(\frac{v_\textrm{wind}}{600 \text{km s}^{-1}}\right)^{-4}
\end{split}
\end{equation}
and Equations \ref{eq:mdot_wind} and \ref{eq:v_wind} can be substituted in for Gaia BH2. We plot the expected X-ray flux from Gaia BH1 and BH2 for a range of possible wind speeds and accretion efficiencies in Figure \ref{fig:xray_est}. 

Figure \ref{fig:xray_est} shows that a 20 ks \textit{Chandra} observation should only be able to detect Gaia BH1 if accretion were radiatively efficient ($\eta \gtrsim 0.01$, depending on wind speed). Due to the high wind speed and low mass loss rate, Gaia BH1 is nowhere near its Eddington luminosity and should not experience radiatively efficient accretion. Indeed, no X-rays are detected from Gaia BH1, which supports this prediction.

Because of the much stronger wind expected for Gaia BH2, it should have approximately 100 times the X-ray flux of Gaia BH1 under the assumption of BHL accretion. This is despite Gaia BH2 being over twice as distant as Gaia BH1. Remarkably, Gaia BH2 would be within the detection threshold of a 20 ks \textit{Chandra} ACIS observation for any values of $\eta_X \gtrsim 10^{-4}$. This also applies for any wind slower than the escape velocity ($\beta_w < 1$), which is to be expected as the wind slows down farther from the star. Figure \ref{fig:xray_est} shows that a 20 ks \textit{Chandra} observation should be able to detect Gaia BH2 down to the case of radiatively \textit{in}efficient flow: $\eta \gtrsim 3\times 10^{-3}$ if the wind speed is the escape velocity and $\eta \gtrsim 10^{-4}$ if the wind slows by the time it escapes from the star and reaches the BH. However, no X-rays were detected from Gaia BH2, indicating that the radiative efficiency is $\eta < 3\times 10^{-3}$. In Figure \ref{fig:xray_est}, we show the expected accretion efficiencies (black dots; assuming BHL accretion rates) using the hot accretion flow models of \cite{2012xie_yuan}, which we will further describe in the following subsection. While these models may not be appropriate for obtaining estimates of accretion efficiency under BHL accretion, the X-ray non-detection of Gaia BH2 shows that reduced \textit{accretion rates}, not just low efficiency at BHL rates, must be invoked to explain this non-detection.

\subsection{Evidence of Reduced Accretion Rate and Inefficient Accretion in Gaia BH2}
The nondetection of X-rays in Gaia BH2 can be explained by going back to Equation \ref{eq:scale_bh1}. The two most uncertain parameters in that equation are the accretion rate at the BH event horizon, $\dot{M}$, as well as the accretion efficiency, $\eta$. Indeed, the former causes a change in the latter \cite[e.g.][]{2012xie_yuan}. We suggest that in Gaia BH2, the X-ray non-detection is due to $\dot{M}$ being lower than the BHL assumption, which also leads to a lower radiative efficiency. This has been seen in other highly sub-Eddington accreting BHs such as the Milky Way's supermassive BH, Sgr A$^*$ \citep[e.g.][]{2003yuan_sgra}, as well as two stellar mass BHs in LMXBs which have been famously well-studied in quiescence: A0620-00 and V404 Cyg \cite[e.g.][]{1996narayan_quiescence, 1997narayan_v404}. 

In all of these systems, a similar reduction in X-rays is seen, and explained by either advection dominated accretion flows (ADAF), or luminous hot accretion flows (LHAF), both of which fall under the class of hot accretion flows \citep[e.g.][]{2012yuan_paper1, 2012xie_yuan}. Most of the energy dissipated by viscosity is stored as entropy rather than being radiated away (e.g. through X-rays). 

Models of hot accretion flows lead to a reduced accretion rate near the BH event horizon, with $\dot{M} \propto r^s$, where $0 < s < 1$. A general description is presented in \cite{2012yuan_paper1}, where it is found that $s \approx 0.5$ and that the accretion rate within $10R_s$ ($R_s$ being the Schwarzschild radius) is approximate constant. This leads to the following reduction to BHL accretion:
\begin{align}
    \dot{M}_\textrm{horizon} = \dot{M}_\textrm{BHL} \left(\frac{10R_s}{R_\textrm{acc}}\right)^{0.5} = \frac{\sqrt{20} v_\textrm{wind}}{c} \dot{M}_\textrm{BHL}
    \label{eq:mdot_correction}
\end{align}
where $R_\textrm{acc} = GM_\textrm{BH}/v_\textrm{w}^2$ is the characteristic radius of accretion. In the case of Gaia BH1, this leads to $\dot{M} \approx 0.009 \dot{M}_\textrm{BHL}$ in Gaia BH1 and $\dot{M} \approx 0.003 \dot{M}_\textrm{BHL}$ in Gaia BH2. 

With a more realistic accretion rate in hand, there is one more correction that we can make, which is to use values of radiative efficiency, $\eta$, computed for hot accretion flows by \cite{2012xie_yuan}. Both accretion rates are highly sub-Eddington: $L_\textrm{Edd} = 0.1\dot{M}_\textrm{Edd}c^2 \Longrightarrow \dot{M}_\textrm{Edd} \approx 2 \times 10^{-7} M_\odot \textrm{ yr}^{-1}$ for both Gaia BH1 and BH2. Gaia BH1 has a predicted accretion rate at the horizon of $\dot{M} \approx 0.009\dot{M}_\textrm{BHL}\approx 5\times 10^{-12}\dot{M}_\textrm{Edd}$ and Gaia BH2 has an accretion rate of $\dot{M} \approx 0.003\dot{M}_\textrm{BHL}\approx 5\times 10^{-9}\dot{M}_\textrm{Edd}$. The same fitting equation is appropriate for both systems:
\begin{align}
    \eta \approx 1.58 \left(100\times\frac{\dot{M}}{\dot{M}_\textrm{Edd}}\right)^{0.65}
    \label{eq:efficiency_correction}
\end{align}
which leads to $\eta\approx 10^{-6}-10^{-5}$ for Gaia BH1 and $\eta\approx 10^{-4}-10^{-3}$ for Gaia BH2 (cyan dots in Figure \ref{fig:xray_est}).

Finally, by substituting both (1) the reduced accretion rate and (2) the corresponding radiative efficiency into Equation \ref{eq:scale_bh1}, we obtain X-ray flux estimates of Gaia BH1 and BH2 to be $10^{-22} - 10^{-20}\,{\rm erg\,s^{-1}\,cm^{-2}} $ and  $10^{-18} - 10^{-16}\,{\rm erg\,s^{-1}\,cm^{-2}} $, respectively, which we show with cyan curves in Figure \ref{fig:xray_est}. This places both systems well under the \textit{Chandra} detection limit, but may be within the limits of future missions.

\subsection{Radio Estimates}

\begin{figure}
    \centering
    \includegraphics[width=0.5\textwidth]{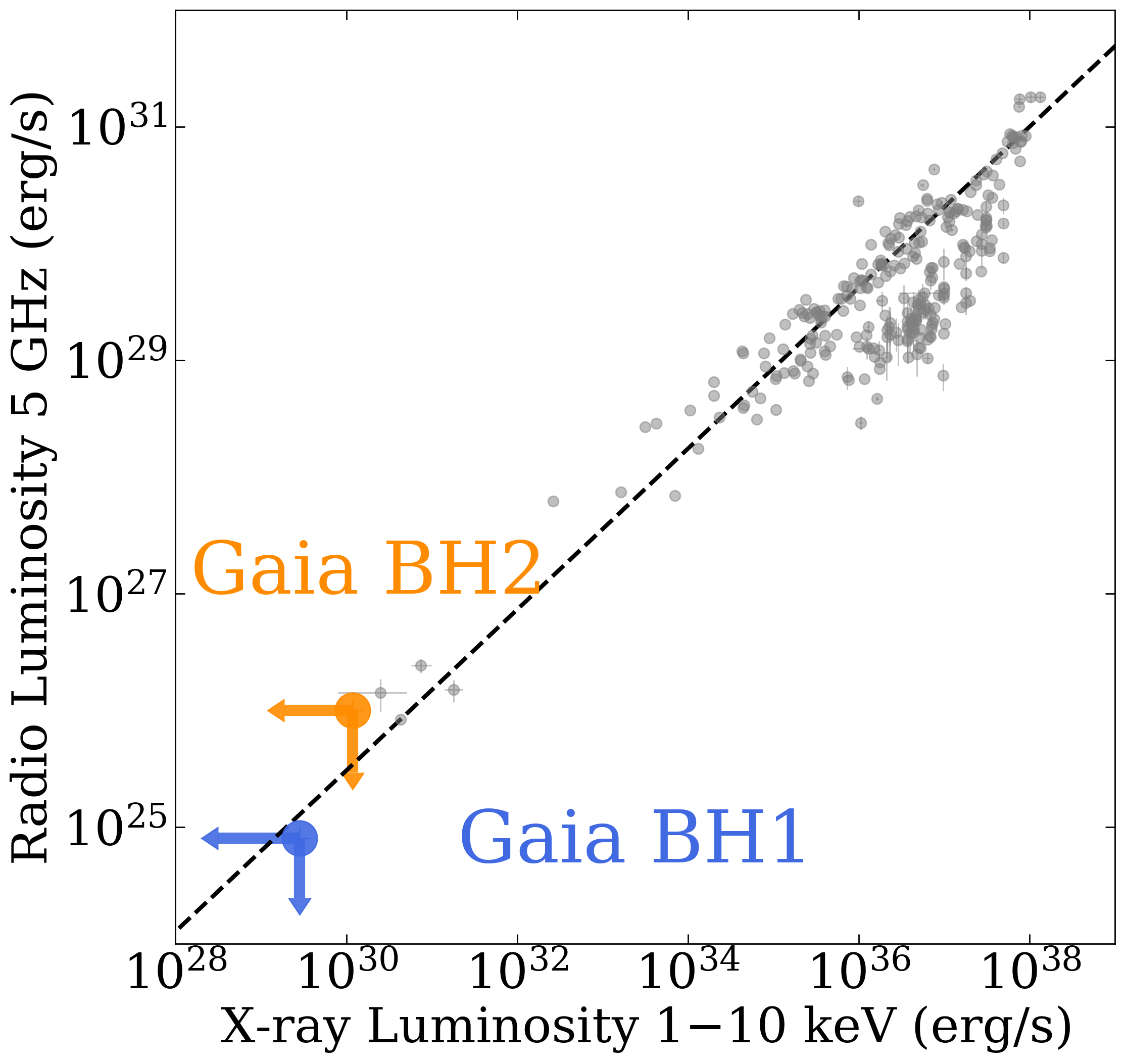}
    \caption{Gaia BH1 and BH2 could lie on the BH ``Fundamental Plane". In gray are all measurements of hard state galactic BHs. The dotted line shows the BH fundamental plane from \cite{2012plotkin} for 10 $M_\odot$ BHs.}
    \label{fig:allBHs}
\end{figure}

The empirical Fundamental Plane of black hole activity relates X-ray luminosity, radio luminosity and BH mass of Galactic black holes and their supermassive analogues \citep{2012plotkin}. By placing BHs on the Fundamental Plane, the physical process behind BH X-ray and radio emission can be understood. We reproduce the most current compilation of hard state Galactic BHs with measured X-ray and radio luminosities \citep{bahramian2018, 2021plotkin} with upper limits of Gaia BH1 and BH2 overplotted in Figure \ref{fig:allBHs}. There is a minor correction to convert to the same X-ray energy ranges and radio frequency ranges, which we omit since it is of order unity.

If we assume that the Fundamental Plane holds for our systems, and our assumptions of reduced accretion rate and inefficient accretion flow, we can calculate the expected radio luminosities/fluxes: $\sim 10^{21} \textrm{erg s}^{-1}$/ $\sim$ 1 nJy at 5 GHz (BH1) and  $\sim 10^{23} \textrm{erg s}^{-1}$/ $\sim$ 10 nJy at 5 GHz (BH2). These radio flux densities are well under the projections for future facilities such as the Next Generation Very Large Array \citep[ngVLA;][]{2018ngvla}, so we proceed with a discussion of prospects for finding other BHs in the X-ray.

\section{Implications for X-ray Searches of BHs}
\label{sec:searches}
If no X-ray or radio signatures of accretion are seen from targeted observations of the two nearest known BHs, then what can we expect from blind searches? In the following subsections, we explore the prospects of detecting in the X-ray, (1) wind-accreting BHs in binary systems similar to Gaia BH2, and (2) isolated BHs accreting from the ISM. While other studies have done similar computations in the past \citep[e.g.][]{2002agol}, we incorporate the modern models of inefficient accretion flow and reduced accretion rates (compared to BHL) which we used to explain the X-ray non-detection of Gaia BH2. 

\subsection{Wind Accreting BHs in Binaries}
Are wind-accreting binaries like Gaia BH2 detectable by current X-ray missions? From Equations \ref{eq:mdot_wind}, \ref{eq:v_wind} and \ref{eq:scale_bh1}, it is clear that systems with (1) a closer separation or (2) a star with a larger radius will lead to a larger X-ray luminosity. 

\begin{figure}
    \centering
    \includegraphics[width=0.5\textwidth]{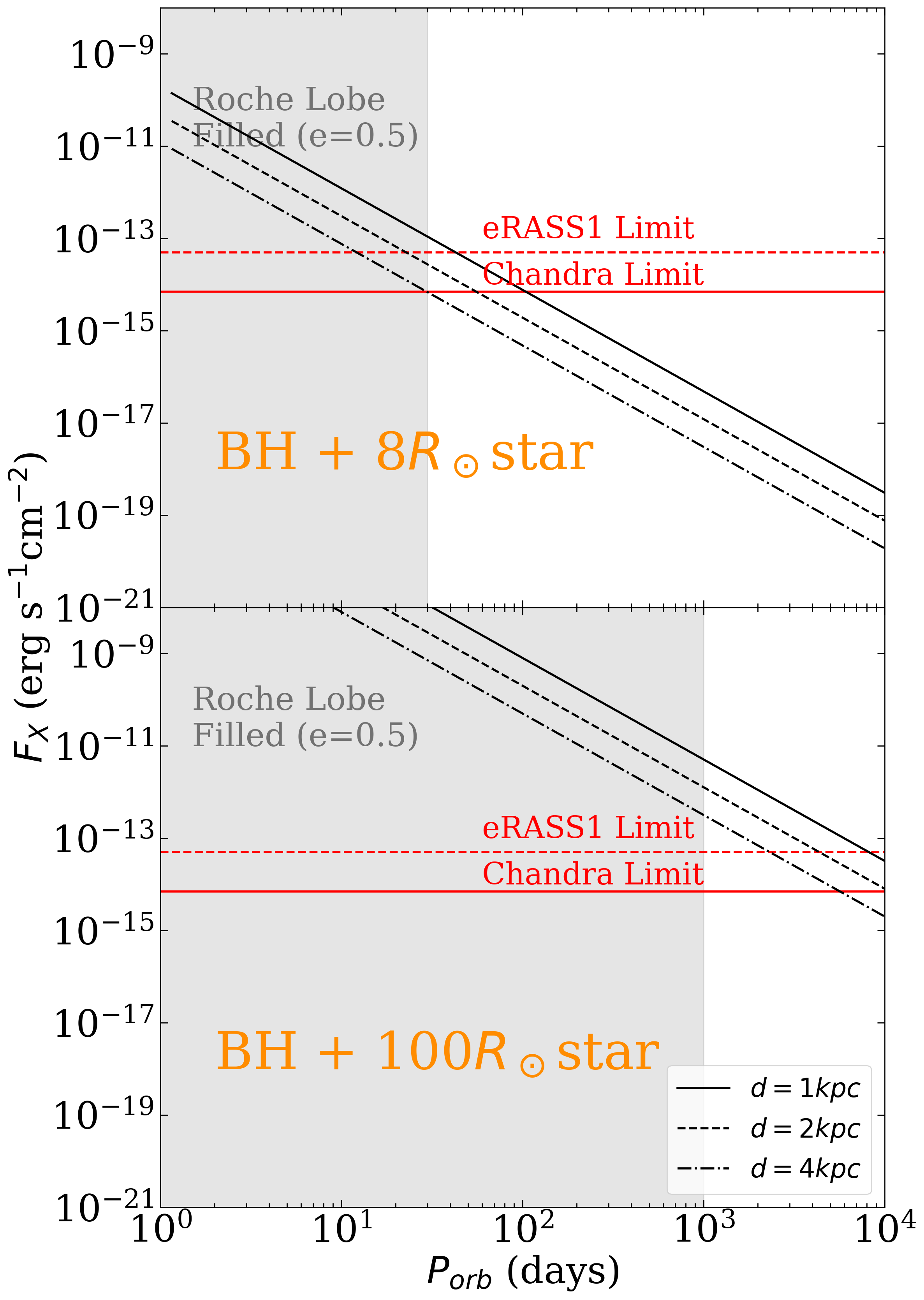}
    \caption{\textit{Top}: Even if Gaia BH2-like system had a shorter orbital period, it would not be detectable in X-rays before filling its Roche lobe (gray shaded area). \textit{Bottom}: A BH in a binary with a tip of the red giant branch star ($R_*\sim 100R_\odot$) can be bright enough in X-rays for the system to be detectable before the star fills its Roche lobe ($P_\textrm{orb} \approx 10^3 - 10^4$ days).}
    \label{fig:binary_est}
\end{figure}

Since sun-like stars that ascend the red giant branch keep their temperatures roughly constant but swell up to $\sim100 R_\odot$, one could expect systems like this to be strong X-ray emitters. In Figure \ref{fig:binary_est}, we show the prospects of finding systems similar to Gaia BH2 from X-ray searches alone. We use Equations \ref{eq:v_wind} and \ref{eq:mdot_wind}, to calculate wind speeds and mass loss rates, and Equations \ref{eq:mdot_correction} and \ref{eq:efficiency_correction} to calculate efficiency and reduced accretion rate corrections from hot accretion flows, as we did for Gaia BH2.
In the top panel of Figure \ref{fig:binary_est}, we plot the X-ray flux as a function of orbital period for a system with all other parameters the same as Gaia BH2 ($M_\textrm{BH} = 9M_\odot$, $R_*=8R_\odot$, $e=0.5$, $d=1.16$ kpc), and observed at periastron. 

Even if a Gaia BH2-like system (i.e. a BH and an $8R_\odot$ giant) were in a shorter period orbit, it would not be detectable before filling its Roche lobe ($P_\textrm{orb}\approx 30$ days). At this point, an accretion disk could form, which could lead to higher radiative efficiency and/or outbursts that could lead to the system being more easily detectable in X-rays. Such calculations are the subject of Section \ref{sec:mesa}, where we explore the prospects of detecting systems similar to Gaia BH1 and BH2 when filling their Roche lobes.

In the bottom panel of Figure \ref{fig:binary_est}, we plot the X-ray flux as a function of orbital period for a system that could resemble what Gaia BH2 will look like in 100 Myr, when the red giant reaches the tip of the red giant branch ($M_\textrm{BH} = 9M_\odot$, $R_*=100R_\odot$, $e=0.5$). We plot the X-ray flux for a system at 1, 2, and 4 kpc. Such a system would fill its Roche lobe at $P_\textrm{orb}\sim 10^3$ days, but systems in the range of $P_\textrm{orb}\sim 10^3 - 10^4$ days are detectable by \textit{Chandra} out to a few kpc, depending on the exact orbital period. In Figure \ref{fig:binary_est} and following figures, we adopt a \textit{Chandra} flux limit of $7\times 10^{-15} \textrm{erg s}^{-1}\textrm{cm}^{-2}$, approximately corresponding to the 3$\sigma$ flux limits presented in this paper in a 20 ks exposure, We also show a flux limit of $5\times 10^{-14} \textrm{erg s}^{-1}\textrm{cm}^{-2}$ from a single all-sky scan of the SRG/eROSITA mission --- eRASS1 is the name of the first all sky scan, though co-adds of multiple scans go deeper \citep{2021sunyaev, 2021predehl}. This means that wind-accreting BHs in binaries could be detectable in X-rays before the donor stars fill their Roche lobes. However, this is only the case for appreciable eccentricities $e \gtrsim 0.5$. Circular orbits (as might be more likely for $R_*\sim 100R_\odot$ donors), would lead to a 75\% decease in flux, pushing the limits of \textit{Chandra}. 

\subsection{BHs Accreting from the ISM}
We compute the observed X-ray flux from a BH accreting from the ISM. This could be either an isolated BH or a BH in a binary or higher-order system, as long as it is accreting from the ISM. Previous works assumed a BHL accretion rate \citep[e.g.][]{2002agol}, whereas we use the corrected accretion rates and efficiencies from \cite{2012yuan_paper1} and \cite{2012xie_yuan}, respectively, as supported by the non-detection of Gaia BH2.

The ISM is made up of at least 5 phases, ordered from most to least dense: gravitationally bound giant molecular clouds (GMCs) made up of molecular hydrogen, diffuse H$_2$ regions, the cold neutral medium (CNM), warm neutral medium (WNM), and warm ionized medium (WIM) \citep[e.g.][]{2011draine}. All phases of the ISM have been found to be roughly in pressure equilibrium (i.e. $\rho \times T = $ constant). Given that sound speed in a medium is proportional to the square root of temperature: $c_s \propto \sqrt{T}$, from Equation \ref{eq:bhl}, it is already clear that ISM-accreting BHs will be more X-ray bright when passing through the densest regions of the ISM.

\begin{figure}
    \centering
    \includegraphics[width=0.49\textwidth]{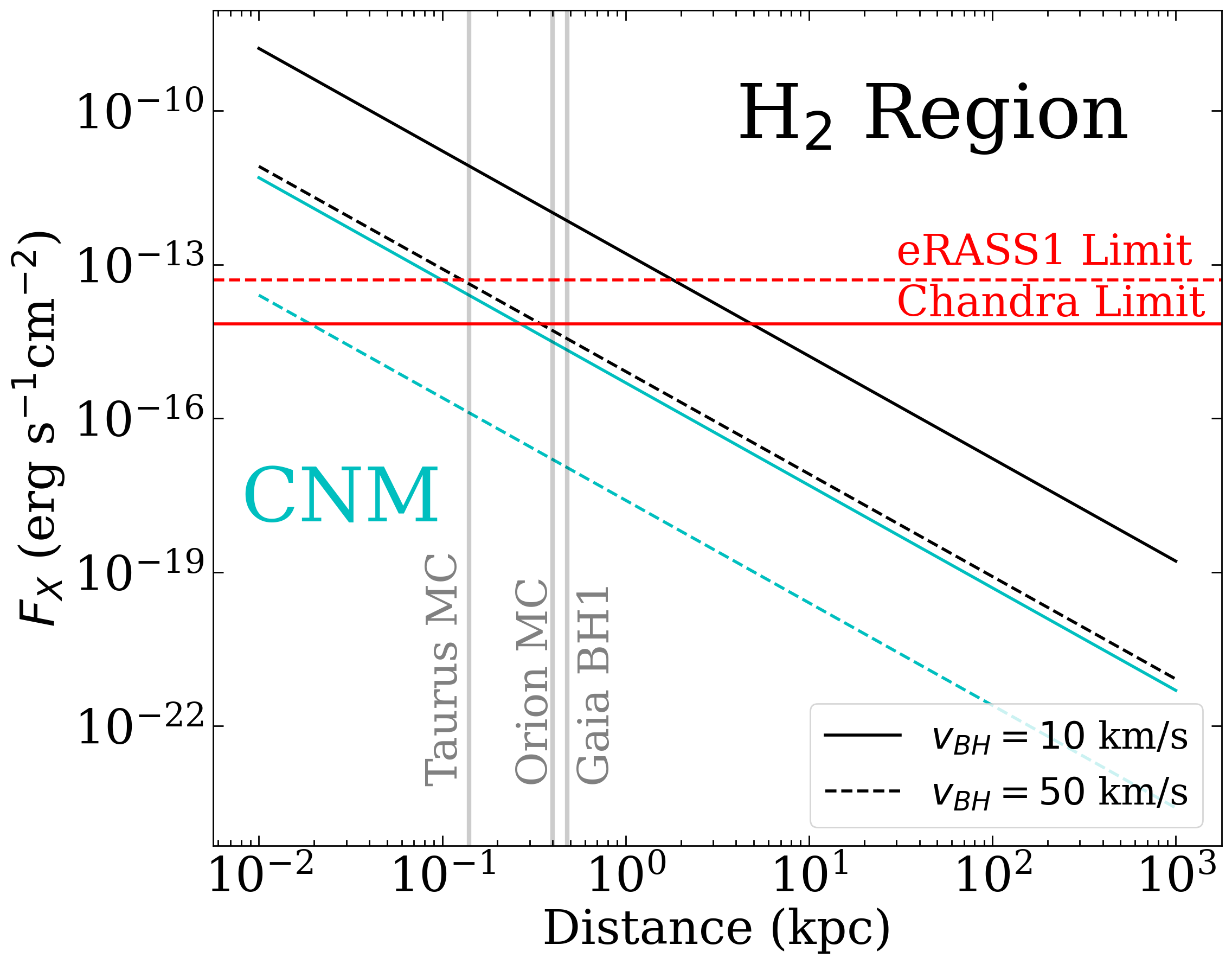}
    \caption{A BH accreting from the ISM in an H$_2$ region is detectable by current X-ray missions out to a few kpc. However, given that the nearest BH, Gaia BH1, is 480 pc away, it is unlikely to find BHs much nearer than that. With that in mind, the plot above shows that the prospects for detecting BHs accreting from the CNM or any lower density ISM phase are slim to none with current capabilities.}
    \label{fig:ism_accretion}
\end{figure}

We calculate the expected X-ray flux due to a BH accreting from an H$_2$ region ($\overline{n} \approx 10^3 \textrm{ cm}^{-3}, T \approx 30$ K) and from the CNM ($\overline{n} \approx 30 \textrm{ cm}^{-3}, T \approx 100$ K). We take $\rho = m_p n$ and calculate the sound speed as $c_s = \sqrt{k_BT/m_p}$. We then use the left hand sides of Equations \ref{eq:bhl} and \ref{eq:fx} and additionally incorporate the hot accretion flow corrections to the accretion rate (Equation \ref{eq:mdot_correction})
and accretion efficiency (Equation \ref{eq:efficiency_correction}).

We plot the X-ray flux as a function of distance for a BH accreting from the ISM in Figure \ref{fig:ism_accretion}. We present curves for a BH accreting from an H$_2$ region and the CNM, for BH space velocities of 5 km/s and 50 km/s. We do not plot curves for higher velocities since the flux levels are reduced dramatically. In other words, isolated BHs with space velocities that exceed 50 km/s are virtually impossible to detect by current X-ray capabilities. While a few isolated BHs within 100 pc {\it may} be detectable as faint X-ray sources, it would be difficult to distinguish them from other astrophysical sources at larger distances.

Figure \ref{fig:ism_accretion} shows that a \textit{very} slow-moving BH ($v = 10$ km/s) can be detectable if passing through a high-density H$_2$ region. Such systems are detectable out to $\sim2$ kpc in eRASS1, and $\sim5$ kpc in a 20 ks \textit{Chandra} pointing. We note that Figure \ref{fig:ism_accretion} makes the propsects of finding such systems deceptively promising, given the low volume fillin factors of H$_2$ regions. 
Furthermore, the high column density of hydrogen in H$_2$ regions is likely to reduce the flux by an appreciable amount, further challenging the prospects of detection.

\begin{figure}
    \centering
    \includegraphics[width=0.49\textwidth]{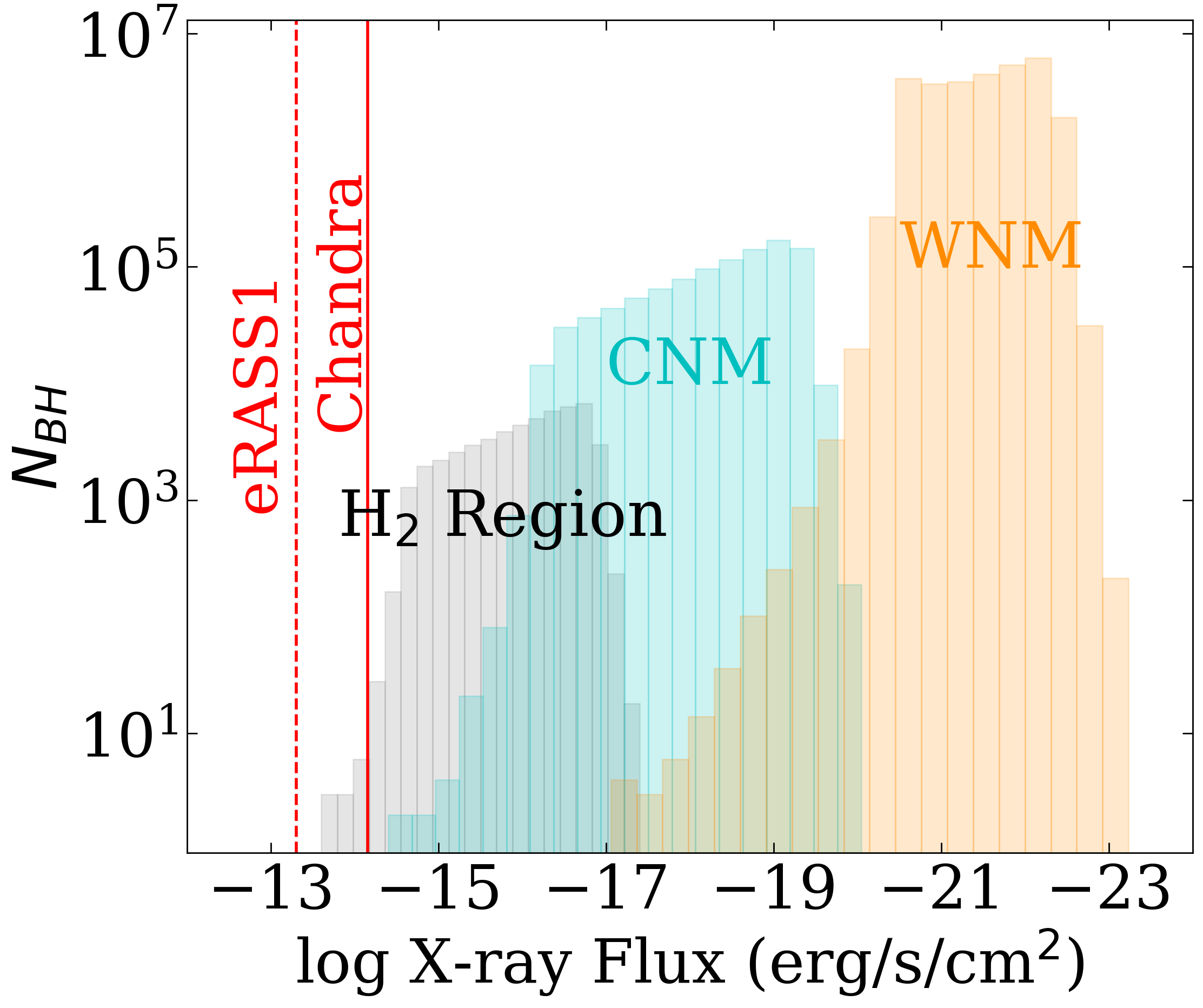}
    \caption{A simple model assuming a distribution of $10^8$ stellar-mass BHs passing through the various phases of the ISM in the Milky Way show that the most X-ray bright will be those passing through low density H$_2$ regions. However, even generous model assumptions suggest that the chances of detecting BHs accreting from the ISM are unlikely after adopting models of hot accretion flows.}
    \label{fig:ism_accretion_pop}
\end{figure}

Given the above calculations for single systems, how many ISM-accreting BHs can be found in the Milky Way? From Equation \ref{eq:bhl} and Figure  \ref{fig:ism_accretion}, it is clear that that $\dot{M}_\textrm{BHL} \propto v_{\rm BH}^{-3}$ scaling relation makes the X-ray flux of BHs dramatically decrease given a slight increase in BH space velocity. Currently, the velocity distribution of BHs in binaries is unknown, both due to low-number statistics (only $\sim20$ systems are dynamically confirmed) \citep{2016corral, 2019atri_bhvelocity} and due to selection effects in samples of detectable BHs. It is  still uncertain if BHs are born with kicks \citep[e.g.][]{2022stevenson_bhkicks, 2023kimball_bhkick}. Furthermore, because there are only a few known BHs in wide binaries and one candidate isolated BH from microlensing, we must assume a velocity distribution.

To investigate an optimistic scenario, we assume the space velocity of ISM-accreting BHs is uniformly distributed between 10 and 50 km/s. We assume that $10^8$ BHs are distributed axisymmetrically throughout the Milky Way, exponentially in cylindrical $h$ and $s$ coordinates with characterstic scales of 410 pc and 1 kpc, respectively \citep[e.g.][]{1995vanparadijs}. We then use the filling factor of each component of the ISM (H$_2$ region: 0.05\%, CNM: 1\%, WMN: 30\%) to calculate the total number of BHs passing through each region \citep[e.g.][]{2011draine}. We present the resulting distributions in Figure \ref{fig:ism_accretion_pop}. Based on those results, virtually no ISM-accreting BHs are detectable in eRASS1, but $\sim10$ could be detectable in 20 ks \textit{Chandra} observations of \textit{all} H$_2$ regions. However, this number is almost certainly inflated due to the effects of a high column density in $H_2$ regions and our uncertain assumptions on the BH velocity distribution. It appears improbable to detect X-rays from ISM-accreting BHs passing through the CNM or WMN, given current capabilities. 

\section{Future evolution of {\it Gaia} Black Holes and Detection as Symbiotic BH XRBs}
\label{sec:mesa}
\begin{figure}
    \centering
    \includegraphics[width=0.5\textwidth]{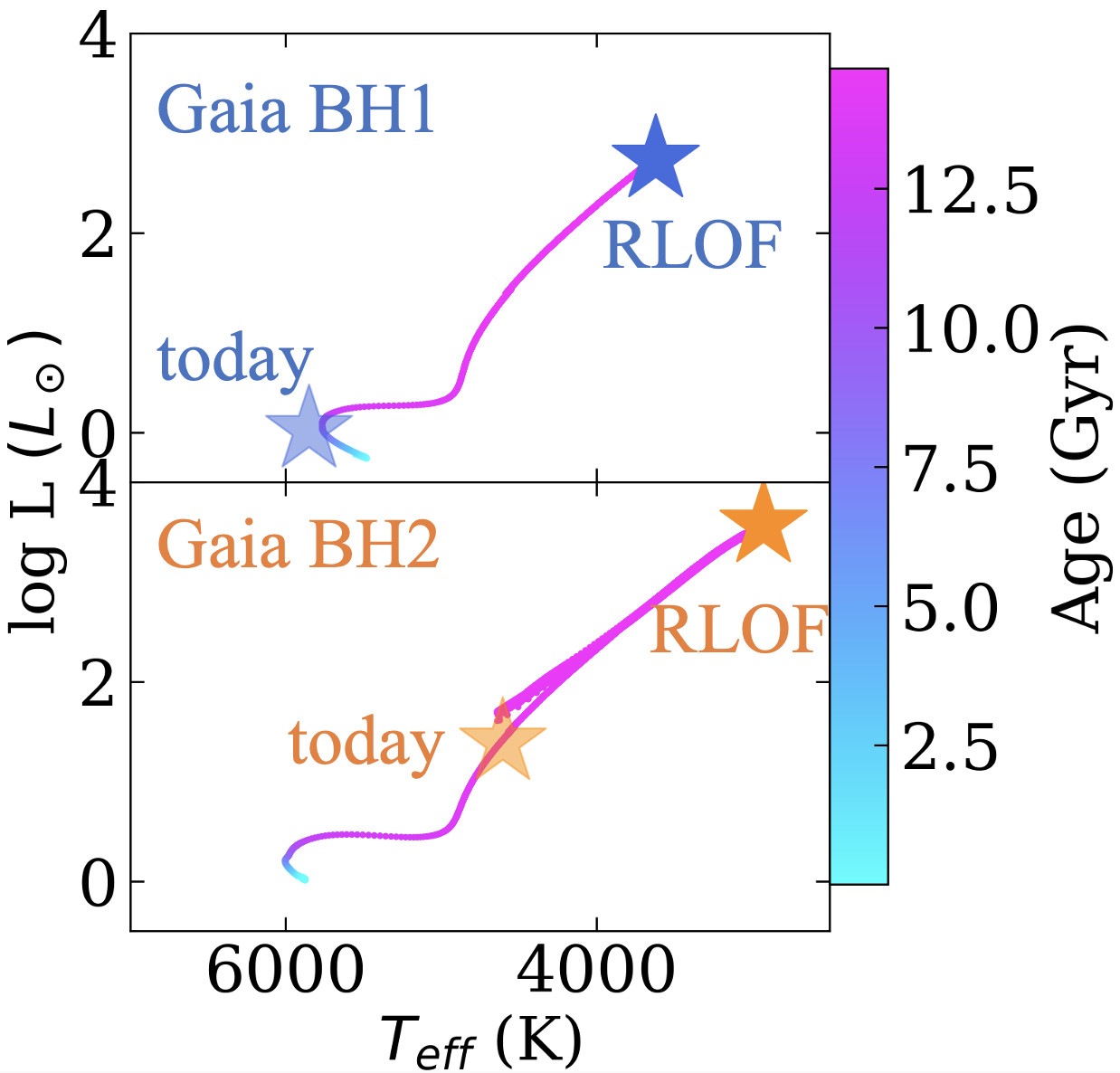}
    \caption{Gaia BH1 will fill its Roche lobe near the tip of its first giant branch in a few Gyr. Gaia BH2 will do so near the tip of the AGB in $\sim100$ Myr. Leading up to this stage, both systems will likely be detectable as symbiotic BH XRBs, yet no such systems have been confirmed to date. }
    \label{fig:mesa_hr}
\end{figure}

We use the expected future evolution of Gaia BH1 and BH2 to understand how common these systems could be in the Milky Way. The discovery paper of Gaia BH1 used the properties of the Gaia DR3 astrometric sample to infer that $\sim40,000$ BH1-like systems should exist \citep[][]{2023bh1}. To constrain the population size, we take a different approach and evolve the Gaia BH1 and BH2 systems using Modules for Experiments in Stellar Astrophysics \citep[MESA;][]{2011mesa, 2013mesa, 2015mesa, 2018mesa}. The donor star in Gaia BH1 will become a red giant in a few Gyr, and will ultimately fill its Roche lobe near the tip of the first giant branch. The donor star in Gaia BH2 is already a red giant, and will swell enough to fill its Roche lobe in $\sim100$ Myr at the tip of the asymptotic giant branch (AGB). We show their locations in the HR diagram today and during RLOF in Figure \ref{fig:mesa_hr}. We look for the timescales in their evolution when the systems could be visible as symbiotic BH XRBs (i.e. a BH accreting from a red giant filling or nearly filling its Roche lobe). 

\begin{figure*}
    \centering
    \includegraphics[width=\textwidth]{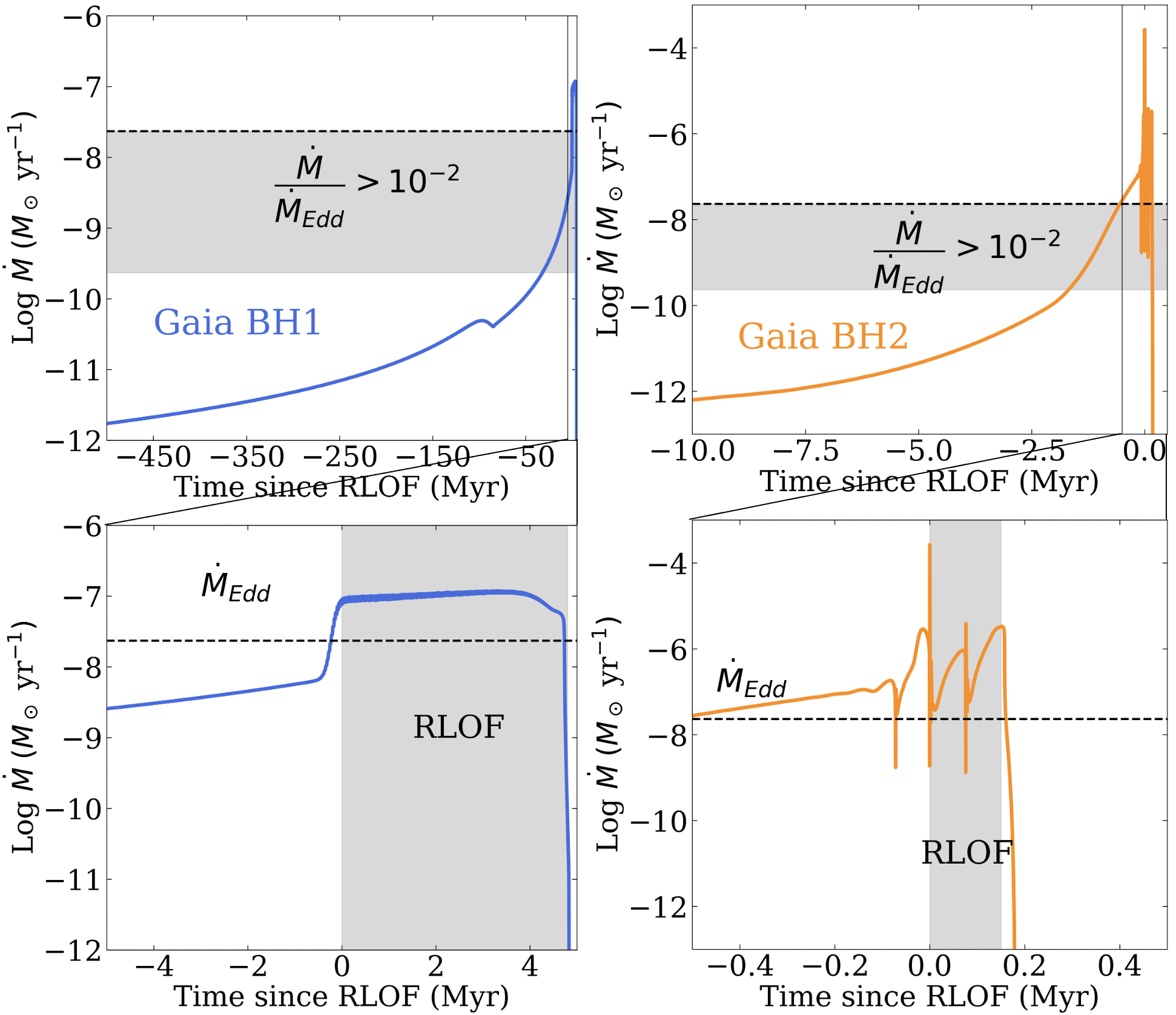}
    \caption{MESA models around the time when the donors fill their Roche lobes show that Gaia BH1 will be visible as a symbiotic BH XRB for $\approx$50 Myr, while $\dot{M} > 10^{-2}\dot{M}_\textrm{Edd}$ (upper left) and for $\approx$5 Myr, while $\dot{M} \sim \dot{M}_\textrm{Edd}$ (lower left). Gaia BH2 will be visible as a symbiotic BH XRB for $\approx$2 Myr, while $\dot{M} > 10^{-2}\dot{M}_\textrm{Edd}$ (upper right) and for $\approx$0.2 Myr, while $\dot{M} \sim \dot{M}_\textrm{Edd}$ (lower right). Because no such systems have been discovered through X-ray bursts, there should be at most $\sim10^4$ Gaia BH1-like systems in the Milky Way, unless outburst timescales of such systems have been underestimated.}
    \label{fig:mesa_bh1}
\end{figure*}

In the top panels of Figure \ref{fig:mesa_bh1}, we show the BH mass loss rates ($\dot{M}$) of the donor star in the Gaia BH1 and BH2 systems for approximately 500 Myr and 10 Myr before Roche lobe overflow (RLOF), respectively. The bottom panels of Figure \ref{fig:mesa_bh1} zoom in and show where RLOF begins. Before RLOF, wind accretion takes place. For low mass loss rates ($\dot{M} \lesssim 10^{-2}\dot{M}_\textrm{Edd, BH}$), the BH accretion rate could be much lower than the donor mass loss rate (as we explain through most of this paper). However, for mass loss rates that approach the Eddington accretion rate of the BH and certainly during RLOF, the two should be nearly equal \citep[e.g.][]{1988ritter}. It is also worth noting that after RLOF, the donor star will have been stripped of its atmosphere, and the orbit of both systems will expand. The orbital period will increase from 186 days to $\approx$850 days in Gaia BH1 and from 1277 days to $\approx$2000 days in Gaia BH2.

In the top panels of Figure \ref{fig:mesa_bh1}, we then shade the region where the accretion rate exceeds $10^{-2}\dot{M}_\textrm{Edd}$, where we expect the accretion rate is high enough to lead to frequent outbursts that could be observed by all-sky X-ray monitors. This accretion rate corresponds to the X-ray lumionsities at which both currently known symbiotic XRBs (albeit with neutron star accretors) have been seen to outburst \citep[e.g.][]{2015kranov_hmxb, 2019yungelson_hmxb}. We define the timescale during which the Gaia BH systems are seen as symbiotic XRBs as $\tau_\textrm{SymXRB}$, which in Gaia BH1 lasts $\approx50$ Myr and in Gaia BH2 lasts $\approx2$ Myr. X-ray monitors such as the \textit{Swift} Burst Alert Trigger (BAT) \citep{2005swift} and MAXI \citep{2009maxi} have similar sensitivities of $\sim$100 mCrab ($10^{-9} \textrm{erg s}^{-1}\textrm{cm}^{-2}$) for $\sim$min long exposures. This means that these monitors are sensitive to essentially all $L_X\sim L_\textrm{Edd}$ bursts in the Galaxy, and sensitive to bursts $L_X\gtrsim 10^{-2}L_\textrm{Edd}$ out to a few kpc. 

Both Gaia BH1 and BH2 undergo a very short phase where $\dot{M} \gtrsim \dot{M}_\textrm{Edd}$ (Gaia BH1 exceeds $\dot{M}_\textrm{Edd}$ by a factor of 10, while BH2 reaches factors of $10^2 - 10^3$). Centaurus X-3 is an example of such a system, where extended periods of low X-ray flux have been observed in a pulsar high mass X-ray binary accreting near the Eddington rate. It has been postulated that at the highest accretion rates, matter gathers at the innermost regions of the accretion disk and absorbs the X-rays, leading to extended lows  \citep{1976cenx3}. However, this system is still X-ray bright for the majority of the time and detectable by all-sky X-ray monitors. Our models show that Gaia BH1 will be in such a phase for $\sim 5$ Myr, and BH2 for $\sim 0.2$ Myr. In both cases, this phase lasts for $\approx$10\% the duration of the evolution when the systems are in the symbiotic XRB phase (i.e. accretion is sub-Eddington and is due to winds rather than Roche lobe overflow). 

In order to estimate an upper limit on the number of similar systems in the Milky Way, we assume a detection efficiency, $\varepsilon_\textrm{detect}$, for all-sky X-ray monitors and take the lifetime of $\sim 1 M_\odot$ stars divided by the time during which these systems are visible as symbiotic BH XRBs:
\begin{equation}
\label{eq:pop_estimate}
\begin{aligned}
&N_\textrm{BH} \sim  \dfrac{\tau_\textrm{stars}}{\varepsilon_\textrm{detect} \tau_\textrm{SymXRB}} 
\\& N_\textrm{BH1}\approx 2\times 10^2 /\varepsilon_\textrm{detect},\quad N_\textrm{BH2}\approx
5 \times 10^3 /\varepsilon_\textrm{detect}
\end{aligned}
\end{equation}

In the case where we use the short-lived phase where the donors are Roche lobe overflowing, the above becomes:
\begin{equation}
\label{eq:pop_estimate_edd}
\begin{aligned}
&N_\textrm{BH} \sim  \dfrac{\tau_\textrm{stars}}{\varepsilon_\textrm{detect} \tau_\textrm{RLOF}} 
\\& N_\textrm{BH1}\approx 2\times 10^3 /\varepsilon_\textrm{detect} ,\quad N_\textrm{BH2}\approx
5 \times 10^4 /\varepsilon_\textrm{detect} 
\end{aligned}
\end{equation}

During the last $\sim$50 years, all-sky X-ray surveys have been sensitive to X-ray bursts from such systems, but no symbiotic BH XRBs have been discovered. From \textit{Uhuru} \citep{1978uhuru} to MAXI, it is highly unlikely that the brightest X-ray bursts have been missed. From Equation \ref{eq:pop_estimate_edd}, even if we assume a 10\% efficiency ($\varepsilon_\textrm{detect}=0.1$) of all-sky X-ray monitors in detecting such systems, this places Gaia BH1-like systems at $N\lesssim 2\times 10^3$ and BH2-like systems at $N\lesssim 2\times 10^4$ in our galaxy. If we assume that systems are only detectable during RLOF, the corresponding limits are $N\lesssim 2\times 10^4$ and $N\lesssim 2\times 10^5$.

There is at least one candidate symbiotic XRB proposed to host a BH, IGR J17454-2919 \citep{2015IGRJ1745}. The most accurate \textit{Chandra} localization of the X-ray source coincides with a red giant (K- to M-type), while the X-ray burst properties of the system do not securely point to either a NS or BH accretor. Ongoing work is being conducted to determine the nature of this system. A handful of symbiotic XRBs hosting NSs have been detected as X-ray sources even though the donors have not yet overflowed their Roche lobes \citep[e.g.][]{Hinkle2006, Masetti2007, Bozzo2018, Hinkle2019, De2022}. That being said, the radiative efficiencies of accreting NSs are likely to be larger than those of BHs \citep[e.g.][]{Garcia2001}, and the accretion rate above which BH symbiotic XRTs are likely to be recognized as such is uncertain.  

\section{Discussion and conclusions}
We have analyzed X-ray and radio observations of the two nearest known BHs: Gaia BH1 and BH2. For both sources, we only detect upper limits in both the X-ray and radio. Due to the relatively strong, low-velocity winds from the red giant in Gaia BH2, BHL accretion predicts that we should have seen X-rays from the system. We interpret our non-detection as a sign of reduced accretion rates as seen in hot accretion flows, and an ensuing lower radiative efficiency than predicted by BHL accretion.  We found that these hot accretion flow corrections lead to X-ray (and radio) fluxes well below the limit of current facilities.


We then used the corrected accretion rates and efficiencies to compute the observed flux from a BH accreting from a $R_*\sim 100R_\odot$ red giant (i.e. what Gaia BH2 will become in 100 Myr). We found that a relatively nearby system ($d\lesssim$4 kpc) of that type could be detectable in X-rays \textit{before} filling its Roche lobe. 

We then extended our calculations to wind-accreting BHs passing through the ISM. We found that the only plausible scenario for detecting such a system would be to have a very slowly moving ($v\lesssim 10$ km/s) BH passing through a dense ($n\gtrsim 10^3\textrm{ cm}^{-3}$) H$_2$ region. Current technologies rule out the possibility of detecting an ISM-accreting BH passing through the CNM or any lower density phase of the ISM, even with generous assumptions about BH velocity and the population distribution.

Finally, we produced MESA models of the future evolution of Gaia BH1 and BH2. We predict that the accretion rate in Gaia BH1 will be high enough ($\dot{M}\sim 10^{-2}\dot{M}_\textrm{Edd}$) for the system to be visible as a symbiotic BH XRB for $\approx$ 50 Myr. The same will be true for Gaia BH2, but only for $\approx$ 2 Myr. Although the symbiotic BH XRB phase is a relatively short-lived phase in the evolution of these systems, the effective search volume for X-ray bright systems is large. Because all-sky X-ray monitors have been sensitive to X-ray bursts in a large part of the Galaxy for the last $\approx$ 50 yrs, the lack of detected symbiotic BH XRBs would seem to imply an upper limit on the number of Gaia BH1-like systems at $N\lesssim 10^3$ ($10^4$ assuming 10\% detection efficiency), assuming BH + giant systems could be detected anywhere in the galaxy when the BH accretes at a rate $\dot{M}> 10^{-2}\dot{M}_\textrm{Edd}$. 

This limit is somewhat puzzling. \citet{2023bh1} estimated that the effective search volume for Gaia BH1-like systems in {\it Gaia} DR3 was only $\sim 3\times 10^6$ stars, which would seem to suggest that $>10^4$ similar systems should exist in the Milky Way. There are several possible explanations for these apparently inconsistent limits.  One is that symbiotic BH XRBs have already been detected by X-ray surveys but have not been recognized as such. This seems plausible particularly for wind-accretion systems, which may not form disks and undergo outbursts. Such systems would appear as relatively faint X-ray sources coincident with red giants. Many such sources exist in the Galactic plane and have never been studied in detail. These consideration suggest that radial velocity follow-up of giants coincident with X-ray sources may be a promising search strategy for symbiotic BH XRBs.  

Another possibility is that the detection efficiency of symbiotic BH XRBs is simply very low. This could be the case if they have unstable disks with very long outburst recurrence timescales, as has indeed been proposed \citep[e.g.][]{Deegan2009}. 

The next few years show promise for the discovery of many more BH binaries: SRG/eROSITA in X-rays, \textit{Gaia} through optical astrometry, and the Rubin Legacy Survey of Space and Time (LSST) through optical photometry. X-ray and radio detections (and non-detections) of future systems will provide new clues regarding the nature of accretion around BHs in a wide range of astrophysical scenarios.

\section{Acknowledgements}
We thank Tim Cunningham for a useful discussion on optimal reduction of \textit{Chandra} data. We are grateful to the \textit{Chandra}, VLA, and MeerKAT directorial offices and support staff for prompt assistance with DDT observations. The MeerKAT telescope is operated by the South African Radio Astronomy Observatory, which is a facility of the National Research Foundation, an agency of the Department of Science and Innovation.

ACR acknowledges support from an NSF Graduate Research Fellowship. ACR thanks the LSSTC Data Science Fellowship Program, which is funded by LSSTC, NSF Cybertraining Grant \#1829740, the Brinson Foundation, and the Moore Foundation; his participation in the program has benefited this work. KE was supported in part by NSF grant AST-2307232.

\bibliography{main}{}
\bibliographystyle{aasjournal}

\end{document}